# Classification of functional fragments by regularized linear classifiers with domain selection

David Kraus* and Marco Stefanucci†

**Abstract:** We consider the problem of classification of functional data into two groups by linear classifiers based on one-dimensional projections of functions. We reformulate the task to find the best classifier as an optimization problem and solve it by regularization techniques, namely the conjugate gradient method with early stopping, the principal component method and the ridge method. We study the empirical version with finite training samples consisting of incomplete functions observed on different subsets of the domain and show that the optimal, possibly zero, misclassification probability can be achieved in the limit along a possibly non-convergent empirical regularization path. Being able to work with fragmentary training data we propose a domain extension and selection procedure that finds the best domain beyond the common observation domain of all curves. In a simulation study we compare the different regularization methods and investigate the performance of domain selection. Our methodology is illustrated on a medical data set, where we observe a substantial improvement of classification accuracy due to domain extension.

**Key words and phrases:** Classification; conjugate gradients; domain selection; functional data; partial observation; regularization; ridge method.

## 1 Introduction

We consider the problem of classification of a functional observation into one of two groups. Classification of functional data is a rich, long-standing topic comprehensively overviewed in Baíllo et al. (2011b). It was recently shown by Delaigle and Hall (2012a) that depending on the relative geometric position of the difference of the group means, representing the signal, and covariance operator, summarizing the structure of the noise, certain classifiers can have zero misclassification probability. This remarkable phenomenon, called perfect classification, is a special property of the infinite-dimensional setting and cannot occur in the multivariate context, unless in degenerate cases. It was demonstrated by Delaigle and Hall (2012a) that a particularly simple class of linear classifiers, based on a carefully chosen one-dimensional projection of the function to

*Department of Mathematics and Statistics, Masaryk University, Kotlářská 2, 611 37 Brno, Czech Republic; *david.kraus@mail.muni.cz*
†Department of Statistical Sciences, Sapienza University of Rome, Piazzale Aldo Moro 5, 00185 Roma, Italy; *marco.stefanucci@uniroma1.it*

classify, can achieve this optimal error rate either exactly or in the limit along a sequence of approximations. Berrendero et al. (2017) further elucidate the perfect classification phenomenon from the point of view of the dichotomy between mutual singularity or absolute continuity between two Gaussian measures on abstract spaces.

Motivated by these findings we reformulate the problem of determining the best classifier as a quadratic optimization problem on a function space, or, equivalently, as a linear inverse problem. These problems are ill-posed which, unlike in most statistical inverse problems, is not a complication but rather an advantage in the sense that the more ill-posed the problem is the better optimal misclassification probability. We use regularization techniques, such as the numerical method of conjugate gradients with early stopping and ridge or Tikhonov regularization, to solve the optimization problem which leads to a class of regularized linear classifiers. From this point of view, the optimal misclassification rate is the limit along the regularization path of solutions which themselves may not converge.

We study the empirical version of the problem, where the objective function in the constrained minimization must be estimated from finite training data. Here our contribution is in two important aspects. First, we show that it is possible to construct an empirical regularization path towards the possibly non-existent unconstrained solution in such a way that the classification error converges to its best, possibly zero, value. We do this specifically for three methods, namely conjugate gradient, principal component and ridge classification, in a truly infinite dimensional manner in the sense that the convergence takes place along a path with decreasing amount of regularization and holds without particular restrictions on the mean difference between classes. Second, all our methodology and theory is developed in the setting of partially observed functional data, where functional trajectories are observed only on subsets of the domain. This type of incomplete data, also called functional fragments, is increasingly common in applications of functional data analysis, e.g., Bugni (2012), Delaigle and Hall (2013), Liebl (2013), Goldberg et al. (2014), Kraus (2015), Delaigle and Hall (2016), Gromenko et al. (2017). Our study is motivated by a medical data set on internal carotid artery aneurysm. The principal difficulty for inference with fragments is that temporal averaging is precluded by the incompleteness of the observed functions. Our formulation as an optimization problem enables to overcome this issue under certain assumptions on the observation pattern because only averaging across individuals in the training data is needed, and not individual curves.

Since the observation domain may vary in the training sample and the new curve to classify also may be observed on a different subset, it is natural to ask which domain should be used for training and application of a classifier. We propose a domain selection strategy that looks for the best classifier with domain ranging from a minimum common domain to the entire domain of the function to classify. For methods of selecting the best observation points previously proposed for different settings and goals we refer to, e.g., Ferraty et al. (2010) and Delaigle et al. (2012).

Our simulation study confirms that this is indeed an effective approach which can result in a considerable reduction of the missclassification probability. Further simula-tions compare the performance of the three types of regularized classifiers. Among other findings, this study shows that the principal component and conjugate gradient classifiers often achieve comparable error rates but the latter usually needs a lower dimension of the regularization subspace, in agreement with a result we provide in the theoretical study.

Application to a data set on the geometric features of the internal carotid artery in patients with and without aneurysm demonstrates the utility of the proposed methodology. These data consist of trajectories observed on intervals of different lengths. Previous analyses of these data used the common domain of all curves in classification. With our results we are able to include information beyond this minimum domain and select the best domain extension, leading to a substantial drop in the error rate of discrimination between risk groups. This gain in efficiency is of course promising also for other applications.

General references on functional data analysis include Ramsay and Silverman (2005), Horváth and Kokoszka (2012) and Hsing and Eubank (2015).

## 2   Regularized linear classification

We regard functional observations as random elements of the separable Hilbert space $L^2(\mathcal{I})$ of square integrable functions on a compact domain $\mathcal{I}$ equipped with inner product $\langle f, g \rangle = \int_{\mathcal{I}} f(t)g(t)dt$ and norm $\|f\| = \langle f, f \rangle^{1/2}$. In most applications $\mathcal{I}$ is an interval and observations are curves but our results can be extended to other objects, such as surfaces or images. We consider classification of a Gaussian random function, $X$, into one of two groups of Gaussian random functions. Group 0 has mean $\mu_0$, group 1 has mean $\mu_1$. Both groups have covariance operator $\mathscr{R}$ defined as the integral operator

$$\mathscr{R}f = \int_{\mathcal{I}} \rho(\cdot, t)f(t)dt$$

with kernel $\rho(s,t) = \text{cov}\{X(s), X(t)\}$. In this section we assume that $\mu_0$, $\mu_1$ and $\mathscr{R}$ are known which corresponds to the asymptotic situation with an infinite training sample.

The function $X$ to classify is Gaussian with unknown mean, either $\mu_0$ or $\mu_1$, and covariance operator $\mathscr{R}$. Like Delaigle and Hall (2012a) we consider the class of centroid classifiers that are based on one-dimensional projections of the form $\langle X, \psi \rangle$, where $\psi$ is a function in $L^2(\mathcal{I})$. If $X$ belongs to group $j$, $j = 0, 1$, the distribution of $\langle X, \psi \rangle$ is normal with mean $\langle \mu_j, \psi \rangle$ and variance $\langle \psi, \mathscr{R}\psi \rangle$. Denote the corresponding Gaussian densities $f_j$. Assuming equal prior probabilities of both classes the optimal classifier assigns $X$ to the class $C(X)$ given by

$$C(X) = 1_{\{f_1(X)/f_0(X) > 1\}} = 1_{\{\langle X - \mu_0, \psi \rangle^2 - \langle X - \mu_1, \psi \rangle^2 > 0\}} = 1_{\{T(X) > 0\}},$$

where $T(X) = \langle X - \bar{\mu}, \psi \rangle \langle \mu, \psi \rangle$ with $\bar{\mu} = (\mu_0 + \mu_1)/2$ and $\mu = \mu_1 - \mu_0$. The misclassifi-cation probability of this classifier is

$$P_0\{C(X) = 1\}/2 + P_1\{C(X) = 0\}/2 = P_0(\langle X - \bar{\mu}, \psi\rangle\langle\mu,\psi\rangle > 0)$$
$$= P_0(\langle X - \mu_0, \psi\rangle > \langle\mu,\psi\rangle/2)$$
$$= 1 - \Phi\left(\frac{\langle\mu,\psi\rangle}{2\langle\psi,\mathscr{R}\psi\rangle^{1/2}}\right).$$

To find the best function $\psi$, one would ideally like to maximize

$$\frac{\langle\mu,\psi\rangle}{\langle\psi,\mathscr{R}\psi\rangle^{1/2}}. \tag{1}$$

Similarly to Delaigle and Hall (2012a) and Berrendero et al. (2017) we see that if $\|\mathscr{R}^{-1}\mu\| < \infty$, then by the Cauchy–Schwarz inequality

$$\frac{|\langle\mu,\psi\rangle|}{\langle\psi,\mathscr{R}\psi\rangle^{1/2}} = \frac{|\langle\mathscr{R}^{-1/2}\mu, \mathscr{R}^{1/2}\psi\rangle|}{\langle\psi,\mathscr{R}\psi\rangle^{1/2}} \leq \frac{\|\mathscr{R}^{-1/2}\mu\|\|\mathscr{R}^{1/2}\psi\|}{\langle\psi,\mathscr{R}\psi\rangle^{1/2}} = \|\mathscr{R}^{-1/2}\mu\|. \tag{2}$$

The equality is achieved for $\psi = \mathscr{R}^{-1}\mu$ for which, or for any positive multiple of it, the probability of misclassification is $1 - \Phi(\|\mathscr{R}^{-1/2}\mu\|/2)$, which is positive due to the finiteness of the quantity $\|\mathscr{R}^{-1/2}\mu\|$ that can be seen as the signal-to-noise ratio.

The maximization of (1) can be solved as the task to

$$\text{maximize } \langle\mu,\psi\rangle \quad \text{subject to } \langle\psi,\mathscr{R}\psi\rangle = 1.$$

Using Lagrange multipliers $\langle\mu,\psi\rangle + \lambda(1 - \langle\psi,\mathscr{R}\psi\rangle)$ and taking Fréchet derivative with respect to $\psi$ one obtains the equation $2\lambda\mathscr{R}\psi = \mu$. Solutions for all $\lambda > 0$, if they exist, i.e., if $\|\mathscr{R}^{-1}\mu\| < \infty$, yield the same optimal misclassification probability. Without loss of generality we take $\lambda = 1/2$. Thus the aim to minimize the error rate translates into the unconstrained quadratic optimization problem to

$$\text{maximize } \langle\mu,\psi\rangle - \tfrac{1}{2}\langle\psi,\mathscr{R}\psi\rangle,$$

or

$$\text{minimize } \tfrac{1}{2}\langle\psi,\mathscr{R}\psi\rangle - \langle\mu,\psi\rangle. \tag{3}$$

If $\psi = \mathscr{R}^{-1}\mu$ does not exist in $L^2(\mathcal{I})$, i.e., $\|\mathscr{R}^{-1}\mu\| = \infty$, there is no maximizer of (1). One can instead consider an approximating, regularized problem that can be solved. Regularization is typically used to solve ill-posed inverse problems, whose solution exists, in a stable way. There, the path of regularized solutions converges to the solution to the problem of interest. Here we are in a different situation in that no solution exists. However, as we will see soon, paths of regularized solutions towards the possibly non-existent solution still turn out to be useful since the misclassification probability converges to the optimal value along these paths.

If a solution exists, one can approximate it by an iterative numerical method. This strategy can be applied also in situations where no solution exists. The idea is to construct a sequence of iterations of an appropriate numerical optimization method. The number of steps taken along this divergent sequence towards the non-existent solution can be seen as a regularization parameter. The conjugate gradient method is particularly suited for this situation.

The first $m$ steps of the conjugate gradient method applied to the linear inverse problem $\mathscr{R}\psi = \mu$, or equivalently to the minimization of the quadratic functional $\frac{1}{2}\langle\psi,\mathscr{R}\psi\rangle - \langle\mu,\psi\rangle$, are described in Algorithm 1. This formulation of the algorithm is based on the multivariate version of Phatak and de Hoog (2002, Section 5) who give further references and also details on how applying the conjugate gradient method to the normal equations in linear regression leads to partial least squares regression. The functions $\nu_j$ are conjugate directions in the sense that $\langle\nu_j,\mathscr{R}\nu_k\rangle = 0$, $j \neq k$, and the functions $\zeta_j$ are called residuals in numerical analysis and are orthogonal, that is, $\langle\zeta_j,\zeta_k\rangle = 0$, $j \neq k$.

**Algorithm 1**. Conjugate gradient regularized classification direction

> Initialize $\psi_0^{\mathrm{CG}} = 0$, $\nu_0 = \zeta_0 = \mu$
> Repeat for $j = 0, \ldots, m-1$
> $\quad f_j = \langle\nu_j,\zeta_j\rangle/\langle\nu_j,\mathscr{R}\nu_j\rangle$
> $\quad \psi_{j+1}^{\mathrm{CG}} = \psi_j^{\mathrm{CG}} + f_j\nu_j$
> $\quad \zeta_{j+1} = \mu - \mathscr{R}\psi_{j+1}^{\mathrm{CG}} \ (= \zeta_j - f_j\mathscr{R}\nu_j)$
> $\quad g_j = -\langle\zeta_{j+1},\mathscr{R}\nu_j\rangle/\langle\nu_j,\mathscr{R}\nu_j\rangle$
> $\quad \nu_{j+1} = \zeta_{j+1} + g_j\nu_j$
> Output $\psi_m^{\mathrm{CG}}$

The conjugate gradient approach is an example of dimension reduction regularization techniques. The method solves the minimization problem (3) with $\psi$ restricted to the Krylov subspace $K_m(\mathscr{R},\mu)$ spanned by $\mu, \mathscr{R}\mu, \ldots, \mathscr{R}^{m-1}\mu$, and also by the first $m$ conjugate directions $\nu_j$ or the first $m$ residuals $\zeta_j$, i.e., it seeks to

$$\text{minimize } \tfrac{1}{2}\langle\psi,\mathscr{R}\psi\rangle - \langle\mu,\psi\rangle \quad \text{subject to } \psi \in K_m(\mathscr{R},\mu).$$

The projection direction that solves this minimization is $\psi_m^{\mathrm{CG}}$.

Another popular choice is to

$$\text{minimize } \tfrac{1}{2}\langle\psi,\mathscr{R}\psi\rangle - \langle\mu,\psi\rangle \quad \text{subject to } \psi \in E_m(\mathscr{R}),$$

where $E_m(\mathscr{R})$ is the subspace spanned by the first $m$ eigenfunctions, $\varphi_1, \ldots, \varphi_m$, of $\mathscr{R}$ in the spectral decomposition

$$\mathscr{R} = \sum_{j=1}^{\infty} \lambda_j \varphi_j \otimes \varphi_j,$$

where $\lambda_1 \geq \lambda_2 \geq \cdots > 0$ are the eigenvalues. The solution $\psi_m^{\mathrm{PC}} = \sum_{j=1}^{m} \lambda_j^{-1}\langle\mu,\varphi_j\rangle\varphi_j$ gives the principal component classifier.

In general one can

$$\text{minimize } \tfrac{1}{2}\langle\psi,\mathscr{R}\psi\rangle - \langle\mu,\psi\rangle \quad \text{subject to } \psi \in S_m,$$

where $S_m$ is the $m$-dimensional subspace generated by some functions $s_1,\ldots,s_m$ such that $s_j$, $j=1,2,\ldots$ generate the range of $\mathscr{R}$. Let $\mathscr{P}_m$ be the projection operator that projects on $S_m$, $\mathscr{R}_m = \mathscr{P}_m\mathscr{R}\mathscr{P}_m$ and $\mathscr{R}_m^- = \mathscr{P}_m\mathscr{R}^{-1}\mathscr{P}_m$. Then the solution of the regularized minimization problem is $\psi_m = \mathscr{R}_m^-\mu$. More explicitly, considering solutions of the form $\psi_m = \sum_{j=1}^m c_j s_j$ leads to the $m$-variate minimization of $\tfrac{1}{2}c^\mathsf{T} Q c - u^\mathsf{T} c$ with the matrix $Q$ with $Q_{jk} = \langle s_j, \mathscr{R} s_k\rangle$ and vector $u$ with $u_j = \langle\mu, s_j\rangle$, i.e., to the solution with coefficients $c = Q^{-1}u$. In the case of the Krylov subspace, the iterative conjugate gradient method outline in Algorithm 1 is, however, preferred because the matrix $Q$ is ill-conditioned.

While $\psi_m$ in general need not converge as $m \to \infty$ since no solution to the unconstrained minimization problem may exist, the misclassification probability associated with the linear classifier given by $\psi_m$ converges along the regularization path.

**Proposition 1.** *The misclassification probability of the regularized linear classifier based on $\psi_m = \mathscr{R}_m^-\mu$ converges to $1 - \Phi(\|\mathscr{R}^{-1/2}\mu\|/2)$ as $m \to \infty$.*

This and all other results are proved in Appendix A.

The above result holds regardless of whether the unconstrained minimization problem (3) has a solution (i.e., regardless of whether $\|\mathscr{R}^{-1}\mu\| < \infty$). The limiting misclassification probability is either positive if $\|\mathscr{R}^{-1/2}\mu\| < \infty$, or zero if $\|\mathscr{R}^{-1/2}\mu\| = \infty$.

The misclassification probability $1 - \Phi(\|\mathscr{R}^{-1/2}\mu\|/2)$ is optimal for classifying between two Gaussian measures with mean difference $\mu$ and covariance $\mathscr{R}$, as explained by Berrendero et al. (2017). It is achieved exactly by the one-dimensional projection on $\psi = \mathscr{R}^{-1}\mu$, when $\|\mathscr{R}^{-1}\mu\| < \infty$. When $\|\mathscr{R}^{-1}\mu\| = \infty$, both dimension reduction techniques, conjugate gradients and principal components, and also ridge regularization introduced later achieve the same optimal limiting misclassification rate along a possibly non-convergent path of one-dimensional projection directions.

It is natural to investigate and compare how quickly the misclassification rate approaches the limit for both main types of subspace regularization. It turns out that the conjugate gradient classifier, being a greedy, goal-oriented procedure, performs better than or at least equally well as the principal component classifier with the same dimension.

**Proposition 2.** *Regardless of whether the optimal misclassification probability can be achieved exactly or along a regularization path, i.e., $\|\mathscr{R}^{-1}\mu\| < \infty$ or $\|\mathscr{R}^{-1}\mu\| = \infty$, and regardless of whether the optimal misclassification probability is zero or positive, i.e., $\|\mathscr{R}^{-1/2}\mu\| = \infty$ or $\|\mathscr{R}^{-1/2}\mu\| < \infty$, the misclassification probability of the principal component classifier using $m$ components is higher than or equal to the misclassification probability of the $m$-step conjugate gradient classifier.*

Principal component and partial least squares regression were previously compared in the multivariate setting by Phatak and de Hoog (2002, Subsection 6.2) who show that "PLS fits closer than PCR." An extension to infinite dimension was provided by Blanchard and Krämer (2010, Theorem 1) who showed in their context of kernel partial least squares that the partial least squares solution is closer to the true solution of the inverse problem than the principal component solution with the same number of components. Unlike these result, our result in Proposition 2 does not assume the existence of a solution of the inverse problem and instead focuses on the values of the functional corresponding to the misclassification probability.

Although Proposition 2 suggests that the conjugate gradient method will typically use less components than the principal component method to achieve the best result, it does not necessarily mean that the resulting misclassification probability with the best number of components will be better for conjugate gradients. We address this question in the simulation study. A similar phenomenon in the context of regression was previously studied in the literature on multivariate partial least squares and recently in the functional setting by Febrero-Bande et al. (2017).

We can also use another approach to regularization based on ridge regression. Optimizing the misclassification probability in a ball with radius $\theta^{1/2}$ leads to the task to

$$\text{minimize } \tfrac{1}{2}\langle \psi, \mathscr{R}\psi \rangle - \langle \mu, \psi \rangle \quad \text{subject to } \|\psi\|^2 \leq \theta$$

which is solved by $\psi_\alpha^{\text{Ridge}} = \mathscr{R}_\alpha^{-1}\mu$, where $\mathscr{R}_\alpha = \mathscr{R} + \alpha\mathscr{I}$, $\alpha \geq 0$ is a regularization parameter and $\mathscr{I}$ is the identity operator. Similarly to the case of subspace regularization we obtain in the following proposition the convergence of the misclassification probability of the ridge classifier to $1 - \Phi(\|\mathscr{R}^{-1/2}\mu\|/2)$ as $\alpha \to 0+$.

**Proposition 3.** *The misclassification probability of the regularized linear classifier based on $\psi_\alpha^{Ridge} = \mathscr{R}_\alpha^{-1}\mu$ converges to $1 - \Phi(\|\mathscr{R}^{-1/2}\mu\|/2)$ as $\alpha \to 0+$.*

Like before, the above result holds regardless of whether the unconstrained minimization problem (3) has a solution (i.e., regardless of whether $\|\mathscr{R}^{-1}\mu\| < \infty$). The limiting misclassification probability is either positive if $\|\mathscr{R}^{-1/2}\mu\| < \infty$, or zero if $\|\mathscr{R}^{-1/2}\mu\| = \infty$.

There is an important difference between the conjugate gradient (partial least squares) method and the other approaches. While principal components and the ridge method regularize the problem without the main goal in mind, the conjugate gradient approach greedily follows the goal of optimal classification. Indeed, the conjugate gradient method as an iterative optimization procedure constructs the regularization path focusing on the minimization of the misclassification probability whereas the other approaches regularize by modifying the operator to be inverted regardless of the objective.

## 3 Empirical classifiers for fragmentary functions

So far we discussed classification assuming that the parameters of each group are known. We now present the empirical version with unknown distributional parameters but with a finite training data set available, and show that such classifiers can achieve the same optimal error rate as if the there were infinite training samples. We aim to do this not only in the case of fully observed functions but also in the case of incomplete curves. Incompleteness can occur in the training data, with each curve possibly observed on a different domain, and in the new curve we wish to classify. One strategy would be to consider all curves on the intersection of their observation domains, if it is non-empty. However, such a restriction can be too severe and is not necessary. We will construct classifiers that use the observed new curve on a set $\mathcal{I}$ which may be its entire observation set or a subset of it without requiring that all training curves be completely observed on $\mathcal{I}$.

For each distribution $P_j$, $j = 0, 1$ with mean $\mu_j$ and covariance operator $\mathscr{R}$ let there be a training sample consisting of $n_j$ curves $X_{j1}, \ldots, X_{jn_j}$. The training data are assumed to be mutually independent. Curves may be observed incompletely with values known only on a subset $O_{ji}$ of the domain and no information about the values on the complement. The observation domains are assumed to be independent of the curves and consist of a finite union of intervals. By $O_{ji}(t)$ we denote the indicator that the $i$th curve in group $j$ is observed at time $t$, that is, $1_{O_{ji}}(t)$; similarly, let $U_{ji}(s,t)$ indicate observation at times $s$ and $t$, i.e., $U_{ji}(s,t) = O_{ji}(s)O_{ji}(t)$.

The mean $\mu_j$ in group $j = 0, 1$ can be estimated by the cross-sectional average

$$\hat{\mu}_j(t) = \frac{1_{[N_j(t)>0]}}{N_j(t)} \sum_{i=1}^{n_j} O_{ji}(t) X_{ji}(t),$$

where $N_j(t) = \sum_{i=1}^{n_j} O_{ji}(t)$ is the total number of observed curves in group $j$ at time $t$. The covariance kernel $\rho(s,t)$ can be estimated by the empirical covariance using pairwise complete observations of groupwise centred curves. Formally, the estimator is

$$\hat{\rho}(s,t) = \frac{M_1(s,t)\hat{\rho}_1(s,t) + M_2(s,t)\hat{\rho}_2(s,t)}{M_1(s,t) + M_2(s,t)},$$

where

$$\hat{\rho}_j(s,t) = \frac{1_{[M_j(s,t)>0]}}{M_j(s,t)} \sum_{i=1}^{n_j} U_{ji}(s,t)\{X_{ji}(s) - \hat{\mu}_{jst}(s)\}\{X_{ji}(t) - \hat{\mu}_{jst}(t)\}$$

and $M_j(s,t) = \sum_{i=1}^{n_j} U_{ji}(s,t)$ and $\hat{\mu}_{jst}(s) = \frac{1_{[M_j(s,t)>0]}}{M_j(s,t)} \sum_{i=1}^{n_j} U_{ji(s,t)} X_{ji}(s)$.

The empirical classifier $\hat{C}(X)$ trained on partially observed curves is defined like the theoretical one with unknown quantities replaced by the estimators introduced above, that is, it assigns a new, independent observation $X$ observed on some domain $\mathcal{I}$ to the class $\hat{C}(X) = 1_{[\hat{T}(X)>0]}$, where $\hat{T}(X) = \langle X - \hat{\bar{\mu}}, \hat{\psi}\rangle\langle\hat{\mu}, \hat{\psi}\rangle$. Here $\hat{\bar{\mu}} = (\hat{\mu}_0 + \hat{\mu}_1)/2$ and $\hat{\mu} = \hat{\mu}_1 - \hat{\mu}_0$ with $\hat{\mu}_j$ being the estimators defined above, possibly restricted to $\mathcal{I}$. The projection direction $\hat{\psi}$ is one of $\hat{\psi}_{m_n}^{\text{CG}}$, $\hat{\psi}_{m_n}^{\text{PC}}$ or $\hat{\psi}_{\alpha_n}^{\text{Ridge}}$ constructed by conjugate gradient, principal component or ridge regularization applied to $\hat{\mu}$ and $\hat{\mathscr{R}}$, with $\hat{\mathscr{R}}$ being the integral operator with kernel $\hat{\rho}(s,t)$ introduced above, possibly restricted to $\mathcal{I} \times \mathcal{I}$.

It is an important feature of all methods discussed in the previous section that they are formulated in terms of the population parameters, i.e., mean difference and covariance operator, and not in terms of individual observations in the training set. The population parameters can be consistently estimated by averaging individual observations whereas temporal averaging of individual curves, e.g., in inner products, is impossible due the incompleteness of the observed functions. In particular, the conjugate gradient method can be applied to fragmentary training data whereas usual algorithms for multivariate or functional partial least squares, e.g., De Jong (1993), Hastie et al. (2009, Algorithm 3.3), Bro and Eldén (2009) and Delaigle and Hall (2012b, Subsection 4.2, Appendix A.2), involve the computation of certain scores, i.e., inner products, for individual curves.

The following assumptions will be needed for the derivation of asymptotic properties of empirically trained regularized linear classifiers.

**Assumption 1.**

(a) *Let the distributions in groups $j = 0, 1$ satisfy $\mathsf{E}_{P_j}(\|X\|^4) < \infty$.*

(b) *For a domain $\mathcal{I}$ let there be $\delta > 0$ such that the observation patterns in training samples $j = 0, 1$ satisfy, as $n_j \to \infty$,*

$$\sup_{(s,t)\in\mathcal{I}\times\mathcal{I}} P\{n_j^{-1} M_j(s,t) > \delta\} = O(n_j^{-2}).$$

Assumption (a) is the standard assumption that guarantees the consistency of the empirical mean and covariance operator for samples of completely observed curves; see, e.g., Bosq (2000) or Horváth and Kokoszka (2012). It was shown in Kraus (2015, Proposition 1) under the additional assumption (b) with $\mathcal{I}$ equal to the entire domain of the curves that the root-$n$ consistency of the sample mean and covariance restricted to $\mathcal{I}$ continues to hold in the fragmentary setting. In particular, it follows that $\|\hat{\mu}_j - \mu_j\| = O_P(n_j^{-1/2})$, and hence $\|\hat{\mu} - \mu\| = O_P(n^{-1/2})$ for $n = \min(n_0, n_1) \to \infty$, and also $\|\hat{\mathscr{R}} - \mathscr{R}\|_\infty = O_P((n_0 + n_1)^{-1/2})$, where $\|\cdot\|_\infty$ is the operator norm. When $\mathcal{I}$ is a subset of the domain, analogous results hold for the obvious restrictions of the functions and integral kernels to $\mathcal{I}$. It is not required that there be any complete curves in the sample, assumption (b) is less restrictive.

We now establish under certain conditions on the regularization path the convergence of the misclassification probability of the empirical conjugate gradient classifier trained on collections of functional fragments to the same optimal limit as for the theoretical conjugate gradient classifier with infinite training sample, regardless of whether the limiting error rate is zero or positive and regardless of whether the limit can be theoretically achieved exactly or along the path.

**Theorem 1.** *Let Assumption 1 hold. Assume that $n = \min(n_0, n_1) \to \infty$ and $m_n \to \infty$ in such a way that $m_n \leq C n^{1/2}$ for some $C > 0$ and*

$$n^{-1/2} \omega_{m_n}^{-1} \|\gamma^{(m_n)}\| + n^{-1} \omega_{m_n}^{-3} \to 0, \tag{4}$$

where $\omega_{m_n}$ is the smallest eigenvalue of the $(m_n \times m_n)$-matrix $H$ with entries $h_{jk} = \langle \kappa_j, \mathscr{R}\kappa_k \rangle$ for $\kappa_j = \mathscr{R}^{j-1}\mu$ and the $m_n$-vector $\gamma^{(m_n)}$ is defined as $\gamma^{(m_n)} = H^{-1}d$ with $d$ being the $m_n$-vector with components $d_j = \langle \mu, \kappa_j \rangle$. Then the misclassification probability of the empirical regularized linear classifier based on $\hat{\psi}_{m_n}^{CG}$ converges in probability to the optimal misclassification probability $1 - \Phi(\|\mathscr{R}^{-1/2}\mu\|/2)$.

The condition in (4) is analogous to (5.10) in Delaigle and Hall (2012b) for partial least squares regression. The vector $\gamma^{(m_n)}$ consists of the coefficients of the theoretical regularized solution $\psi_{m_n}^{\mathrm{CG}}$ with respect to the non-orthogonal basis $\kappa_1, \ldots, \kappa_{m_n}$ of the Krylov subspace $K_{m_n}(\mathscr{R}, \mu)$, i.e., $\psi_{m_n} = \sum_{j=1}^{m_n} \gamma_j^{(m_n)} \kappa_j$. The eigenvalues of $H$ are called the Ritz values in numerical analysis; for details in connection with partial least squares see Lingjærde and Christophersen (2000).

In the proof in Subsection A.4 we make use of the results of Delaigle and Hall (2012b) on the consistency of partial least squares regression for functional data. These results were obtained for situations that are different from our setting in several ways. In particular, we work with functional fragments instead of complete curves, the conjugate gradient path differs from partial least squares regression, e.g., in the group centring in the estimation of the covariance, and we do not require that the population inverse problem, $\mathscr{R}\psi = \mu$ in our context, have a solution. However, our inspection of the underlying technical arguments in Delaigle and Hall (2012b) showed that appropriate analogous results can be obtained and used in our setting, as we explain in the proof.

Next, we show that the empirically trained principal component classifier with increasing number of components asymptotically achieves the optimal misclassification probability.

**Theorem 2.** *Let Assumption 1 hold. Assume that $n = \min(n_0, n_1) \to \infty$ and $m_n \to \infty$ in such a way that $\lambda_{m_n}^4 n \to \infty$ and*

$$\frac{\lambda_{m_n}^2 n}{(\sum_{j=1}^{m_n} a_j)^2} \to \infty,$$

*where $a_1 = 2^{3/2}(\lambda_1 - \lambda_2)^{-1}$ and $a_j = 2^{3/2}\max\{(\lambda_{j-1} - \lambda_j)^{-1}, (\lambda_j - \lambda_{j+1})^{-1}\}$, $j = 2, 3, \ldots$ Then the misclassification probability of the empirical regularized linear classifier based on $\hat{\psi}_{m_n}^{PC}$ converges in probability to the optimal misclassification probability $1 - \Phi(\|\mathscr{R}^{-1/2}\mu\|/2)$.*

The conditions on the principal component regularization path are the same as in the case of functional principal component regression (Cardot et al., 1999). Unlike in the functional linear model it is not assumed that the inverse problem has a solution since the goal is not to estimate the possibly non-existent bounded linear regression functional. For the asymptotic study of the misclassification probability it is enough to show that the empirical and theoretical linear functional approach each other which is guaranteed by the conditions of the theorem.

Finally, the empirical ridge classifier with finite training data asymptotically attains the same optimal error rate as its theoretical counterpart.

**Theorem 3.** *Let Assumption 1 hold. Assume that $n = \min(n_0, n_1) \to \infty$ and $\alpha_n \to 0$ in such a way that $\alpha_n^4 n \to \infty$. Then the misclassification probability of the empirical regularized linear classifier based on $\hat{\psi}_{\alpha_n}^{Ridge}$ converges in probability to the optimal misclassification probability $1 - \Phi(\|\mathscr{R}^{-1/2}\mu\|/2)$.*

Unlike with conjugate gradients and principal components, the conditions on the ridge regularization path do not involve parameters of the data generating distributions because no subspace is constructed.

The regularization parameter can be selected by minimizing an estimate of the misclassification probability. We use leave-one-out cross-validation. When the data contain incomplete curves, the cross-validation procedure must run over the set of complete curves only, i.e., only splits in which the test curve left out is complete can be considered. The incomplete curves cannot be left out because the classifier cannot be applied to them and, therefore, it is necessary to assume for cross-validation that there are enough complete curves in the data on the considered domain, although in the estimation of the classifier itself no complete curves are needed. The best value of the regularization parameter is searched for over a grid of values, e.g., the values corresponding to integer degrees of freedom up to some maximum value. The degrees of freedom for the subspace methods are the dimension of the subspace and for the ridge method they are defined as the trace of $(\hat{\mathscr{R}} + \alpha\mathscr{I})^{-1}\hat{\mathscr{R}}$. The maximum number of degrees of freedom we use is one fifth of the number of curves.

We conclude this section by comparing our results with previous results on near-perfect classification, incomplete functional data and related topics. Delaigle and Hall (2012a) explore the near-perfect classification phenomenon focussing on the theoretical linear classifier regularized by subspace methods. The consistency of the empirical version is established in Delaigle and Hall (2013) for the principal component linear and quadratic classifier based on partially observed training data. Berrendero et al. (2017) work in the setting of complete curves and use dimension reduction regularization by evaluation of functional observations at a finite set of arguments. They show the consistency of the empirical version but do not study asymptotics for decreasing amount of regularization, i.e., do not let the dimension grow. Baíllo et al. (2011a) study optimal classifiers for Gaussian measures based Radon–Nikodym derivatives and investigate the performance of their empirical version in the special class of processes with triangular covariance functions. In contrast, all our methods, including the ridge approach not considered previously, are developed for fragmentary training samples and shown to achieve the optimal misclassification rate for general Gaussian processes along the empirical regularization path. One-dimensional projections of functional data are the basis of other inferential methods, e.g., goodness-of-fit tests (Cuesta-Albertos et al., 2007). Cuevas (2014) provides a comprehensive overview.

## 4 Domain selection

As mentioned at the beginning of the previous section, in the presence of incomplete curves one may restrict attention to the intersection of the observation domains of all curves, say $\mathcal{I}_0$, and apply regularized linear classification or any other existing functional data classification method to the set of complete, restricted curves. An obvious drawback of this approach is that one can possibly lose discrimination power because the difference between the classes may be more pronounced outside $\mathcal{I}_0$. One of the main advantages of our methodology discussed above is its capability to work with incomplete curves since the empirical construction of the projection direction only requires the estimation of $\mu$ and $\mathscr{R}$ which is possible in the presence of missing data. Hence one may look at a larger domain than $\mathcal{I}_0$. A natural choice is the largest subset of the observation set of the function to classify containing enough data for the estimation of the classifier, i.e., satisfying Assumption 1(b), and enough complete functions for leave-one-out cross-validation. This way one uses the largest possible region of argument values hoping to capture the widest range shapes of the group difference. On the other hand, not even this maximal domain, say $\mathcal{I}^{\max}$, may lead to the best classification accuracy because one includes more uncertainty in the estimation due to missing values and moreover the mean difference may not be important in the added part of the domain. Therefore, it seems reasonable to look also at intermediate choices of the domain between the two extremes, $\mathcal{I}_0$ and $\mathcal{I}^{\max}$.

We propose a domain selection strategy for choosing the best interval for the most common case of interval observation sets. The idea is to construct the classifier on a series of intervals, ranging from the common domain $\mathcal{I}_0$ to the maximal domain $\mathcal{I}^{\max}$, extending step by step the working interval by a fixed percentage. More formally, we consider a sequence of nested intervals $\mathcal{I}_0 \subset \mathcal{I}_1 \subset \cdots \subset \mathcal{I}_k \subset \cdots \subset \mathcal{I}_K = \mathcal{I}^{\max}$ starting from $\mathcal{I}_0$ and ending in $\mathcal{I}_K = \mathcal{I}^{\max}$ and on each of them build the linear classifier. The regularization parameter for the $k$th classifier is selected by cross-validation in which we leave out only the units that are fully observed in the working domain $\mathcal{I}_k$. Finally, among these $K+1$ candidates we select the one that minimizes the cross-validation estimate of error.

This search strategy can be extended by considering larger systems of candidate domains, e.g., one can vary both endpoints separately. Also, the idea of domain selection can be generalized to other situations, e.g., non-interval observation sets, multivariate functional data with components with the same or different argument variables, or functions indexed by multivariate arguments. In each situation on needs to define a meaningful system of domains between the common and maximal domain and optimize the cross-validation score over it.

## 5 Simulations

### 5.1 Behaviour of regularized classifiers on complete data

In this section we illustrate the behaviour of the three estimators of $\psi$ under different settings. In particular, we consider Gaussian processes with covariance kernel $\rho(s,t) = \exp(-|s-t|^2/0.01)$ and mean function depending on the group label. Group 0 has mean $\mu_0(t) = 0$ in each setting. Group 1 has mean $\mu_1(t) = \mu(t)$ for we we consider eight different forms. Settings (i)–(iii) correspond to linear, quadratic and cubic trends, setting (iv) consists of a sinusoidal wave, settings (v) and (vi) are exactly the first and the tenth eigenfunction of $\mathscr{R}$ and settings (vii) and (viii) are, respectively, a symmetric and an asymmetric beta density on $\mathcal{I} = [0, 1]$. Mathematical formulas of these functions can be found in Appendix B.

In each of 5000 repetitions we generated 50 curves from group 0 and 50 curves from group 1 and evaluated them on a grid of 100 equispaced points in $\mathcal{I} = [0, 1]$. We also generated a new observation that could arise from group 0 or group 1 with equal probability. Then we constructed the regularized classification direction by the principal component, conjugate gradient and ridge method with $m$ degrees of freedom and predicted the label of the new observation. We considered $m = 1, \ldots, 20$, corresponding to a reasonable minimum of five observations per degree of freedom.

The results are plotted in Fig. 1. It shows the misclassification proportion over the 5000 repetitions as a function of $m$ for different choices of $\mu(t)$. As expected, the conjugate gradient method performs well in all settings and is not much affected by the particular shape of the mean difference. By contrast, the performance of the principal component classifier is strongly dependent on $\mu(t)$. To see this, consider two extreme situations in settings (v) and (vi). The classification error of principal components is close to the one of conjugate gradients in case (v), where $\mu(t)$ is the first eigenfunction, but it is much higher at lower dimensions in case (vi), where $\mu(t)$ is the tenth eigenfunction. In the latter case, the principal component method reaches the same level of error as the conjugate gradient method only when $m = 10$ or more. These findings agree with the theoretical result of Proposition 2 and also with conclusions of Delaigle and Hall (2012a) and Febrero-Bande et al. (2017) who point out that principal components need more degrees of freedom than partial least squares to reach good performance. In this regard ridge regularization seems to be between the two subspace methodologies. It is more similar to conjugate gradients in most cases. In particular in case (vi) it does not completely fail at low degrees of freedom because it does not construct a subspace that can possibly miss the important information. On the other hand it also suffers in this situation, where $\mu(t)$ is on the tail of the spectrum, because ridge penalization shrinks higher index spectral components more than lower index components. However, with sufficiently many degrees of freedom differences fade away and the three methods behave similarly.

## 5.2 Performance of cross-validation for selection of degrees of freedom

We employed simulations to investigate the performance of leave-one-out cross validation in choosing the right level of regularization. The settings were the same as before but classification was done using the number of degrees of freedom selected by leave-one-out cross-validation. We summarize the classification error in Table 1. A general observation is that cross-validation performs well as a selector of the best amount of regularization since the value of misclassification rate in Table 1 is in each case close to the corresponding minimum error in Fig. 1. Principal components appear to perform worst while the conjugate gradient and ridge methods have comparable performance. Table 2 reports

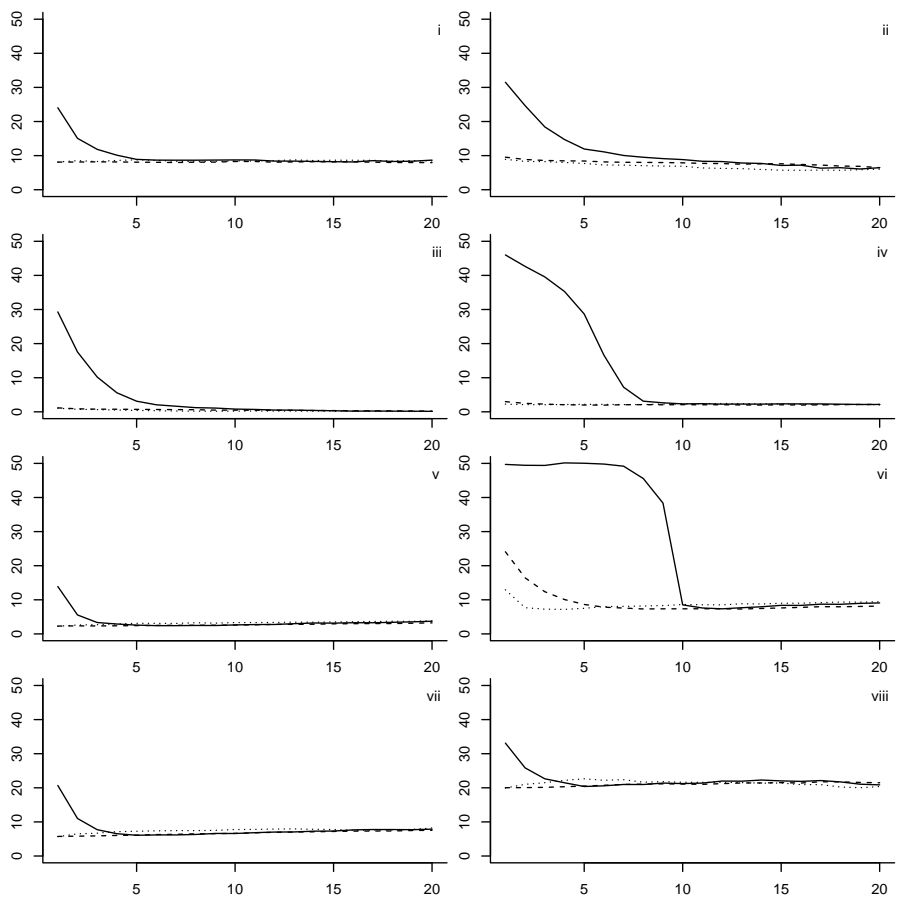

Figure 1: Misclassification rate (%) versus degrees of freedom for different forms of $\mu(t)$, (i) linear, (ii) quadratic, (iii) cubic, (iv) sinusoidal, (v) first eigenfunction, (vi) tenth eigenfunction, (vii) symmetric beta, (viii) asymmetric beta, for principal component (solid), conjugate gradient (dotted) and ridge (dashed) classifiers.

Table 1: Misclassification rate (%) and its standard error for classifiers with degrees of freedom selected by cross-validation for different settings

|    | (i) | (ii) | (iii) | (iv) | (v) | (vi) | (vii) | (viii) |
|----|-----|------|-------|------|-----|------|-------|--------|
| PC | 13.0 (0.34) | 8.3 (0.28) | 1.3 (0.11) | 2.5 (0.16) | 7.2 (0.26) | 7.6 (0.27) | 10.7 (0.31) | 26.2 (0.44) |
| CG | 8.6 (0.28) | 6.5 (0.25) | 0.7 (0.09) | 2.1 (0.14) | 2.6 (0.16) | 7.8 (0.27) | 6.1 (0.24) | 20.9 (0.41) |
| R  | 8.4 (0.28) | 7.7 (0.27) | 0.7 (0.09) | 2.2 (0.15) | 2.4 (0.15) | 7.9 (0.27) | 6.1 (0.24) | 20.8 (0.41) |

PC, principal components; CG, conjugate gradients; R, ridge.

Table 2: Mean (and median) degrees of freedom selected by cross-validation for different settings

|    | (i) | (ii) | (iii) | (iv) | (v) | (vi) | (vii) | (viii) |
|----|-----|------|-------|------|-----|------|-------|--------|
| PC | 8.2 (7) | 14.3 (15) | 9.9 (9) | 10.9 (10) | 4.6 (4) | 11.9 (11) | 5.3 (4) | 8.6 (6) |
| CG | 5.4 (3) | 10.7 (11) | 3.4 (2) | 4.5 (2) | 2.4 (1) | 4.9 (3) | 2.7 (1) | 8.6 (7) |
| R  | 6.4 (3) | 11.6 (13) | 6.0 (3) | 6.1 (4) | 2.7 (1) | 9.3 (8) | 3.4 (1) | 6.7 (3) |

PC, principal components; CG, conjugate gradients; R, ridge.

the mean and median selected degrees of freedom. We see that the principal component method often uses considerably more degrees of freedom than the other methods. This is particularly interesting in case (v), where the mean difference equals the first eigenfunction and thus one component should be the best choice in theory. These results once again document the general phenomenon that principal components are not appropriate for inference about means due to the possible lack of informativeness of the principal components about the mean and the extra uncertainty associated with the estimation of these components.

### 5.3 Missing data and domain selection

We now show the usefulness of the domain selection methodology presented in Section 4. We considered Gaussian processes on $[0, 1]$ with the same covariance as before and with three scenarios for the mean difference of the form of a multiple of a beta density. These were the (i) Beta(2,6), (ii) Beta(5,5) and (iii) Beta(6,2) density which reflect situations in which discrimination due to a peak is in the left, central or right part of the domain, respectively. We sampled 50 curves from group 0 and 50 curves from group 1 on a sequence of 100 equispaced points in $[0, 1]$. Then we generated endpoints of the observation interval for each curve from the uniform distribution on $(0.5, 1)$, that is, each curve was observed between 0 and the endpoint and missing beyond the endpoint. Also the new observation had an endpoint sampled between 0.5 and 1. So the first half of $[0, 1]$, $\mathcal{I}_0 = [0, 0.5]$, was the common observation domain of all curves. We considered extensions of $\mathcal{I}_0$ to $\mathcal{I}_k = [0, 0.5 + 0.05k]$, $k = 0, \ldots, 9$. For each interval of this form that was contained in the observation domain of the curve to classify we estimated the classifiers choosing the best degrees of freedom via cross-validation and classified the new curve. This was repeated 1000 times. We show the behaviour of the resulting

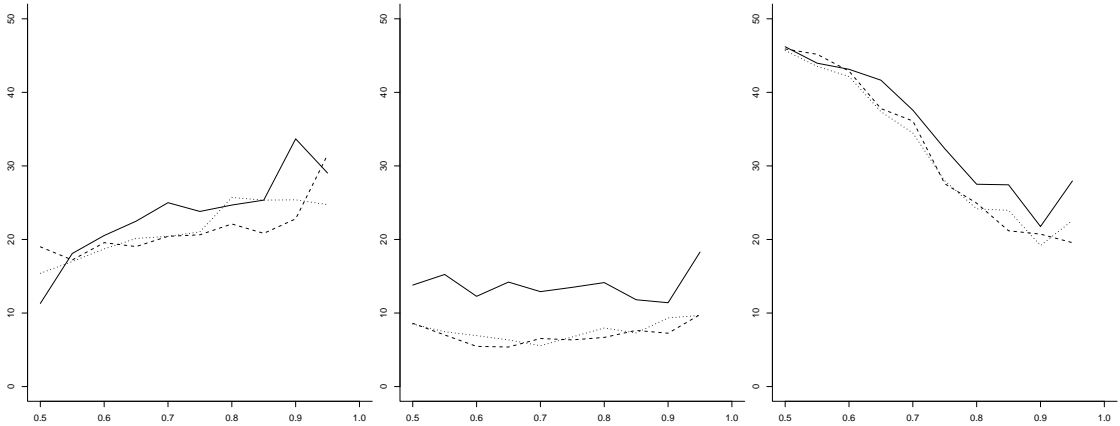

Figure 2: Misclassification rate (%) as a function of the domain extension for $\mu(t)$ being the Beta(2,6) (left), Beta(5,5) (middle), Beta(6,2) (right) density for principal component (solid), conjugate gradient (dotted) and ridge (dashed) classifiers with selected degrees of freedom. Classification is performed on the domains $[0, u]$, $u \in [0.5, 0.95]$, error values are plotted against $u$.

classification error and as a function of the endpoint of the extended domain in Fig. 2.

In the case, where the peak of the mean difference is in the left part of $[0, 1]$, extending the domain does not lead to better classification. The reason is that in this case the interval, where discrimination is, corresponds to the part of the domain where all the data are available and inflating the domain only incurs uncertainty due to the presence of missing data. In the second case the peak of the mean difference is exactly at 0.5 and extending the domain leads to little improvement. The last case is the converse of the first one, the discrimination is mainly in the right part of $[0, 1]$. Here extending the domain considerably reduces the classification error because good classification is only possible by employing the right part of the domain. The classification error is about 45 % using only $\mathcal{I}_0$ but drops to about 20 % using also part of the interval where the data are partially observed.

We see that domain extension may or may not lead to an improvement of the performance of classifiers, depending on the interplay between the form of the mean difference, the covariance structure and the missingness pattern. Practically, one selects the best extension by cross-validation.

# 6 AneuRisk data example

We apply the proposed methodology to the AneuRisk dataset from an interdisciplinary project that aimed at investigating the role of vessel morphology, blood fluid dynamics and biomechanical properties of the vascular wall on the pathogenesis of cerebral aneurysms. See Sangalli et al. (2014b) for an introduction to the data. This dataset has been previously analyzed in several works with different methodological focuses, such as function and derivative estimation (Sangalli et al., 2009b), exploratory analysis and classification (Sangalli et al., 2009a), alignment and clustering (Sangalli et al., 2014a) among others.

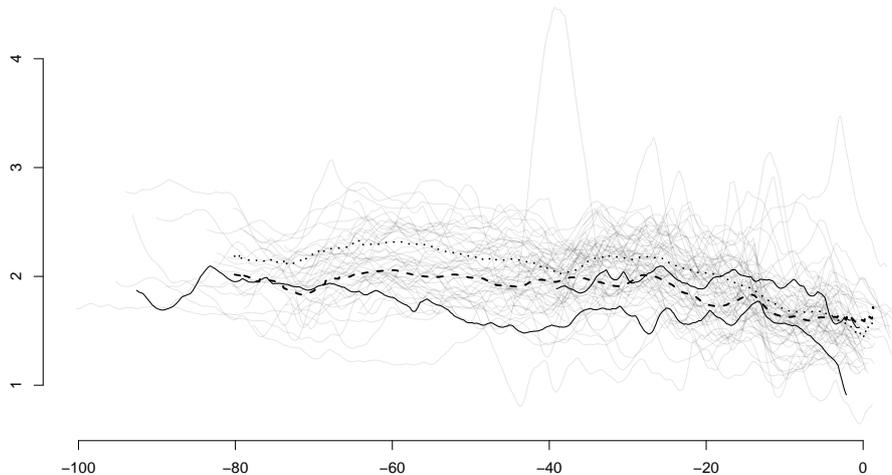

Figure 3: Radius curves in the AneuRisk dataset, along with the mean for the group with an aneurysm after the bifurcation (dotted) and before the bifurcation or without aneurysm (dashed). Curves for two example subjects are highlighted by solid lines. Note the different start and end points for different subjects in the study.

The data consist of measurements of the radius and curvature of the internal carotid artery in a sample of 65 patients of which 33 have an aneurysm at the bifurcation of the vessel or after it, while the other 32 have an aneurysm before the bifurcation, which is a much less dangerous condition, or are healthy. The goal is to classify the patients using the morphology of their internal carotid artery. In this illustration we work with one of the observed variables, the radius. The data have previously been preprocessed, registered and smoothed and are observed on a grid of 2000 points in $[-100.3, 5.1]$, where the argument represents the distance between the observation point and the terminal bifurcation of the internal carotid artery, with positive values for points inside the skull. As we can see in Figure 3, the data are partially observed because the start and end points are different from subject to subject. All subjects are observed on the subset $\mathcal{I}_0 = [-32.9, -7.4]$ that corresponds to $24.3\,\%$ of the whole domain.

We first apply the regularized linear classifiers to curves restricted to the common domain $\mathcal{I}_0$. The classification error estimated by cross-validation is $29.2\,\%$ for principal component, $29.2\,\%$ for conjugate gradient and $32.3\,\%$ for ridge regularized classification.

We compare this procedure with a different approach consisting of a multivariate classification method applied to principal component scores. Specifically, the covari-ance kernel is estimated from observations centred to their respective group means, its eigenfunctions are computed and quadratic discriminant analysis is applied to the inner products of the uncentred curves with the eigefunctions. This procedure is similar to that in Sangalli et al. (2009a). The best classifier of this type turns out to exhibit a misclassification error of 32.3 %, obtained with 2 eigenfunctions.

These values show that in this data set, when attention is restricted to the common domain $\mathcal{I}_0$, the proposed methodology is comparable to the more standard multivariate technique.

Next, we consider classification on extended domains including observed values outside the common domain $\mathcal{I}_0$. We build the sequence of domains $\mathcal{I}_0, \ldots, \mathcal{I}_K$ by enlarging the domain at each step by 1.25 % of the complement of $\mathcal{I}_0$. This step size is a compromise between the fineness of the grid and the computational cost. We consider extended domains up to $K = 40$, corresponding to $\mathcal{I}_{40} = [-66.6, -1.2]$, because not enough subjects have observed values outside this interval for reliable estimation and cross-validation. All regularized linear classification methods obtained a benefit from the domain extension, in particular the error rate for principal components dropped from 29.2 % to 23.2 %, for conjugate gradients from 29.2 % to 25.8 % and for ridge regularization from 32.3 % to 25 %. The best domain was $\mathcal{I}_{10} = [-41.3, -5.8]$ for the conjugate gradient method and $\mathcal{I}_{11} = [-42.2, -5.7]$ for the other two methods.

It should be noted that the alternative method consisting of multivariate classification in the space of scores cannot be applied on extended domains since the individual scores of incomplete curves cannot be computed, although they can be predicted (Kraus, 2015). By constrast, the proposed methods are entirely formulated in terms of distributional parameters, which can be consistently estimated from incomplete data, unlike individual quantites.

## Acknowledgement

The AneuRisk data and useful comments were kindly provided by Laura Sangalli. The work of David Kraus was supported by the Czech Science Foundation (project GJ17-22950Y).

## A  Proofs

### A.1  Proof of Proposition 1

The misclassification probability for $\psi_m$ is
$$1 - \Phi\left(\frac{\langle \mu, \psi_m \rangle}{2\langle \psi_m, \mathscr{R}\psi_m \rangle^{1/2}}\right).$$

Using the fact that $\psi_m \in S_m$, we compute
$$\frac{\langle \mu, \psi_m \rangle}{\langle \psi_m, \mathscr{R}\psi_m \rangle^{1/2}} = \frac{\langle \mu, \mathscr{R}_m^- \mu \rangle}{\langle \mu, \mathscr{R}_m^- \mathscr{R} \mathscr{R}_m^- \mu \rangle^{1/2}} = \|(\mathscr{R}_m^-)^{1/2} \mu\|.$$

The right-hand side in the equation above converges by Lebesgue's monotone convergence theorem to $\|\mathscr{R}^{-1/2}\mu\|$ (finite or infinite), and, therefore, the limiting misclassification probability that is attained along the regularization path $\psi_m$, $m \to \infty$ is $1 - \Phi(\|\mathscr{R}^{-1/2}\mu\|/2)$.

## A.2 Proof of Proposition 2

The conjugate gradient method minimizes the quadratic objective function in the Krylov subspace $K_m(\mathscr{R}, \mu)$ whose elements are in the form

$$\eta = \sum_{k=0}^{m-1} c_k \mathscr{R}^k \mu = p(\mathscr{R})\mu,$$

where $p$ is a polynomial of order lower than $m$. Then $\eta \in K_m(\mathscr{R}, \mu)$ can be written as

$$\eta = \sum_{j=1}^{\infty} p(\lambda_j) b_j \varphi_j$$

with $b_j = \langle \mu, \varphi_j \rangle$. The objective function at $\eta$ equals

$$\begin{aligned}
\tfrac{1}{2}\langle \eta, \mathscr{R}\eta \rangle - \langle \mu, \eta \rangle &= \tfrac{1}{2}\langle p(\mathscr{R})\mu, \mathscr{R} p(\mathscr{R})\mu \rangle - \langle \mu, p(\mathscr{R})\mu \rangle \\
&= \sum_{j=1}^{\infty} b_j^2 \{\tfrac{1}{2} p(\lambda_j)^2 \lambda_j - p(\lambda_j)\} \\
&= \sum_{j=1}^{\infty} \frac{b_j^2}{2\lambda_j} q(\lambda_j)\{q(\lambda_j) - 2\},
\end{aligned} \tag{5}$$

where $q(\lambda) = p(\lambda)\lambda$ is a polynomial of degree at most $m$ such that $q(0) = 0$. The conjugate gradient method finds the polynomial with these properties that minimizes the objective function. We shall find a polynomial $q$ with the required properties such that the objective function above is smaller than or equal to the objective function for the principal component classifier; this will complete the proof.

The principal component classifier uses

$$\psi_m^{\text{PC}} = \sum_{j=1}^{m} \frac{b_j}{\lambda_j} \varphi_j,$$

the objective function at $\psi_m^{\text{PC}}$ equals

$$\tfrac{1}{2}\langle \psi_m^{\text{PC}}, \mathscr{R}\psi_m^{\text{PC}} \rangle - \langle \mu, \psi_m^{\text{PC}} \rangle = -\sum_{j=1}^{m} \frac{b_j^2}{2\lambda_j}. \tag{6}$$

Consider the polynomial of degree $m$ with $q(0) = 0$ given by

$$q(\lambda) = 1 - (-1)^m \frac{\lambda - \lambda_1}{\lambda_1} \cdots \frac{\lambda - \lambda_m}{\lambda_m}.$$

We see that $q(\lambda_j) = 1$ for $j = 1, \ldots, m$ and so the first $m$ summands in the series (5) and (6) are equal. For $j > m$ it holds that $0 \leq q(\lambda_j) \leq 2$ due to the properties of the eigenvalue sequence, thus $q(\lambda_j)\{q(\lambda_j) - 2\} \leq 0$, and, therefore, the corresponding summands in the series (5) are negative whereas they are zero in the series (6). Hence for this polynomial

$$\sum_{j=1}^{\infty} \frac{b_j^2}{2\lambda_j} q(\lambda_i)\{q(\lambda_i) - 2\} \leq -\sum_{j=1}^{m} \frac{b_j^2}{2\lambda_j}$$

and so the objective at the conjugate gradient solution must be smaller than or equal to the objective at the principal component solution.

The inequality between the minima of the quadratic objective function implies the inequality between the misclassification probabilities stated in the proposition.

### A.3 Proof of Proposition 3

Proceeding like in the proof of Proposition 1 we need to show that

$$\frac{\langle \mu, \mathscr{R}_\alpha^{-1} \mu \rangle}{\langle \mu, \mathscr{R}_\alpha^{-1} \mathscr{R} \mathscr{R}_\alpha^{-1} \mu \rangle^{1/2}} = \frac{\sum_{j=1}^{\infty} \frac{b_j^2}{\lambda_j + \alpha}}{\left(\sum_{j=1}^{\infty} \frac{\lambda_j b_j^2}{(\lambda_j + \alpha)^2}\right)^{1/2}} \xrightarrow{\alpha \to 0+} \left(\sum_{j=1}^{\infty} \frac{b_j^2}{\lambda_j}\right)^{1/2} = \|\mathscr{R}^{-1/2} \mu\|,$$

where $b_j = \langle \mu, \varphi_j \rangle$ is the coefficient of $\mu$ in the eigenbasis. If $\sum_{j=1}^{\infty} b_j^2/\lambda_j < \infty$, the convergence follows from Lebesgue's monotone convergence theorem. Otherwise, we use the inequality $\sum_{j=1}^{\infty} \lambda_j b_j^2/(\lambda_j + \alpha)^2 \leq \sum_{j=1}^{\infty} b_j^2/(\lambda_j + \alpha)$ to bound the expression on the left-hand side from below by $\sum_{j=1}^{\infty} b_j^2/(\lambda_j + \alpha)$ which diverges to infinity by Lebesgue's monotone convergence theorem. This completes the proof.

### A.4 Proof of Theorem 1

The probability of misclassifying a new observation using the conjugate gradient classifier based on $\hat{\psi}_{m_n}^{\mathrm{CG}}$ is

$$1 - \Phi\left(\frac{\langle \mu, \hat{\psi}_{m_n}^{\mathrm{CG}} \rangle}{2\langle \hat{\psi}_{m_n}^{\mathrm{CG}}, \mathscr{R} \hat{\psi}_{m_n}^{\mathrm{CG}} \rangle^{1/2}}\right). \tag{7}$$

We need to show that the fraction above converges in probability to $\|\mathscr{R}^{-1/2}\mu\|/2$ along the regularization path satisfying the assumptions of the theorem.

To deal with the numerator in (7) we rewrite it as

$$\langle \mu, \hat{\psi}_{m_n}^{\mathrm{CG}} \rangle = (\langle \mu, \hat{\psi}_{m_n}^{\mathrm{CG}} \rangle - \langle \mu, \psi_{m_n}^{\mathrm{CG}} \rangle) + \langle \mu, \psi_{m_n}^{\mathrm{CG}} \rangle.$$

It can be shown that

$$\langle \mu, \hat{\psi}_{m_n}^{\mathrm{CG}} \rangle - \langle \mu, \psi_{m_n}^{\mathrm{CG}} \rangle = O_P(n^{-1/2} \omega_{m_n}^{-1} \|\gamma^{(m_n)}\| + n^{-1} \omega_{m_n}^{-3}).$$

This result follows from an analog of (5.9) in Theorem 5.3 in Delaigle and Hall (2012b) (and intermediate results in the proof of that theorem) which can be established in our context. Specifically, the necessary modifications of the proofs of Theorems 5.1, 5.2 and 5.3 in Delaigle and Hall (2012b) are as follows. All results remain valid for incomplete instead of complete curves because the proofs depend only on the root-$n$ consistency of the covariance estimators which is satisfied for functional fragments as well (see Proposition 1 of Kraus, 2015). Moreover, derivations in Delaigle and Hall (2012b) can be repeated without assuming that the theoretical solution $\psi = \mathscr{R}^{-1}\mu$ exists as an element of the $L^2(I)$ space; indeed, the proofs in Delaigle and Hall (2012b) are based on stochastic expansions of $\hat{\mathscr{R}}^j \psi = \hat{\mathscr{R}}^j \mathscr{R}^{-1}\mu$ (in our notation) around $\mathscr{R}^j \psi = \mathscr{R}^j \mathscr{R}^{-1}\mu = \mathscr{R}^{j-1}\mu$ and derived quantities but the same steps can be done for $\hat{\mathscr{R}}^{j-1}\hat{\mu}$ around $\mathscr{R}^{j-1}\mu$ present in our situation. In other words, it holds that $\hat{\psi}_{m_n}^{\mathrm{CG}}$ and $\psi_{m_n}^{\mathrm{CG}}$, the empirical and theoretical regularized solution, converge to each other without assuming that $\psi_{m_n}^{\mathrm{CG}}$ converges.

Similarly, for the denominator in (7) we write

$$\langle \hat{\psi}_{m_n}^{\mathrm{CG}}, \mathscr{R}\hat{\psi}_{m_n}^{\mathrm{CG}} \rangle = (\langle \hat{\psi}_{m_n}^{\mathrm{CG}}, \mathscr{R}\hat{\psi}_{m_n}^{\mathrm{CG}} \rangle - \langle \psi_{m_n}^{\mathrm{CG}}, \mathscr{R}\psi_{m_n}^{\mathrm{CG}} \rangle) + \langle \psi_{m_n}^{\mathrm{CG}}, \mathscr{R}\psi_{m_n}^{\mathrm{CG}} \rangle.$$

It holds that

$$\langle \hat{\psi}_{m_n}^{\mathrm{CG}}, \mathscr{R}\hat{\psi}_{m_n}^{\mathrm{CG}} \rangle - \langle \psi_{m_n}^{\mathrm{CG}}, \mathscr{R}\psi_{m_n}^{\mathrm{CG}} \rangle = O_P(n^{-1/2}\omega_{m_n}^{-1}\|\gamma^{(m_n)}\| + n^{-1}\omega_{m_n}^{-3}).$$

This last result is analogous to (7.27) of Delaigle and Hall (2012b) whose proof can be repeated with the same modifications for our situation as before.

Therefore, regardless of whether $\|\mathscr{R}^{-1}\mu\|$ (or $\|\mathscr{R}^{-1/2}\mu\|$) is finite or infinite, we see that the theoretical and empirical regularized quantities approach each other and the misclassification probability is

$$1 - \Phi\left(\frac{\langle \mu, \psi_{m_n}^{\mathrm{CG}} \rangle + o_P(1)}{2\langle \psi_{m_n}^{\mathrm{CG}}, \mathscr{R}\psi_{m_n}^{\mathrm{CG}} \rangle + o_P(1)}\right).$$

The required result follows like in the proof of Proposition 1.

## A.5 Proof of Theorem 2

Similarly to the proof of Theorem 1 we show that

$$1 - \Phi\left(\frac{\langle \mu, \hat{\psi}_{m_n}^{\mathrm{PC}} \rangle}{2\langle \hat{\psi}_{m_n}^{\mathrm{PC}}, \mathscr{R}\hat{\psi}_{m_n}^{\mathrm{PC}} \rangle^{1/2}}\right) \tag{8}$$

converges in probability to $1 - \Phi(\|\mathscr{R}^{-1/2}\|/2)$. The strategy of the proof is similar to that for Theorem 3.1 of Cardot et al. (1999) for the principal component approach to the functional linear model. The difference is in the incompleteness of the functional observations and in that we do not assume that the underlying theoretical inverse problem has a solution.

We rewrite

$$\|\hat{\psi}_{m_n}^{\mathrm{PC}} - \psi_{m_n}^{\mathrm{PC}}\| \leq \|\hat{\mathscr{R}}_{m_n}^- - \mathscr{R}_{m_n}^-\|_\infty \|\hat{\mu}\| + \|\mathscr{R}_{m_n}^-\|_\infty \|\hat{\mu} - \mu\|.$$

Proceeding like in the proof of Lemma 5.1 in Cardot et al. (1999) we can show that

$$\|\hat{\mathscr{R}}_{m_n}^- - \mathscr{R}_{m_n}^-\|_\infty \leq \frac{\|\hat{\mathscr{R}} - \mathscr{R}\|_\infty}{\hat{\lambda}_{m_n} \lambda_{m_n}} + \frac{2\|\hat{\mathscr{R}} - \mathscr{R}\|_\infty \sum_{j=1}^{m_n} a_j}{\lambda_{m_n}}.$$

Here $\hat{\lambda}_j$ are the eigenvalues of $\hat{\mathscr{R}}$ in descending order and $\hat{\varphi}_j$ are the corresponding eigenfunctions. When establishing the above inequality one uses the facts that $|\hat{\lambda}_j - \lambda_j| \leq \|\hat{\mathscr{R}} - \mathscr{R}\|_\infty$ and $\|\hat{\varphi}_j - \text{sign}\langle\hat{\varphi}_j, \varphi_j\rangle\varphi_j\| \leq a_j\|\hat{\mathscr{R}} - \mathscr{R}\|_\infty$ which are known from Bosq (2000, Lemmas 4.2 and 4.3) for the empirical covariance operator from complete curves but hold also for functional fragments (see the proof of Proposition 2 in the supplementary document for Kraus, 2015). Combining this with the facts that $\|\hat{\mu}\| = O_P(1)$, $\|\mathscr{R}_{m_n}^-\| = \lambda_{m_n}^{-1}$, $\|\hat{\mu} - \mu\| = O_P(n^{-1/2})$ and $\|\hat{\mathscr{R}} - \mathscr{R}\|_\infty = O_P(n^{-1/2})$ gives

$$\|\hat{\psi}_{m_n}^{\text{PC}} - \psi_{m_n}^{\text{PC}}\| \leq \lambda_{m_n}^{-2} O_P(n^{-1/2}) + \lambda_{m_n}^{-1} O_P(n^{-1/2}) \sum_{j=1}^{m_n} a_j$$

on the event $[\hat{\lambda}_{m_n} > \hat{\lambda}_{m_n}/2]$; the probability of the complementary event is bounded by $\lambda_{m_n}^{-2} O_P(n^{-1})$. Therefore, $\|\hat{\psi}_{m_n}^{\text{PC}} - \psi_{m_n}^{\text{PC}}\| \to 0$ under the assumptions of the theorem.

Thus $\langle\mu, \hat{\psi}_{m_n}^{\text{PC}}\rangle - \langle\mu, \psi_{m_n}^{\text{PC}}\rangle \to 0$ and $\langle\hat{\psi}_{m_n}^{\text{PC}}, \mathscr{R}\hat{\psi}_{m_n}^{\text{PC}}\rangle - \langle\psi_{m_n}^{\text{PC}}, \mathscr{R}\psi_{m_n}^{\text{PC}}\rangle \to 0$, that is, the estimation error due to the finite training sample vanishes. Consequently, the asymptotic behaviour of the misclassification probability is driven by the behaviour of the theoretical classifier addressed in Proposition 1.

### A.6  Proof of Theorem 3

We show that the fraction in the argument of $\Phi$ in the misclassification probability

$$1 - \Phi\left(\frac{\langle\mu, \hat{\psi}_{\alpha_n}^{\text{Ridge}}\rangle}{2\langle\hat{\psi}_{\alpha_n}^{\text{Ridge}}, \mathscr{R}\hat{\psi}_{\alpha_n}^{\text{Ridge}}\rangle^{1/2}}\right). \tag{9}$$

converges in probability to $\|\mathscr{R}^{-1/2}\mu\|/2$ as $n \to \infty$.

Rewrite the numerator as

$$\langle\mu, \hat{\psi}_{\alpha_n}^{\text{Ridge}}\rangle = \langle\mu, \hat{\mathscr{R}}_{\alpha_n}^{-1}\hat{\mu}\rangle = \langle\mu, (\hat{\mathscr{R}}_{\alpha_n}^{-1} - \mathscr{R}_{\alpha_n}^{-1})\hat{\mu}\rangle + \langle\mu, \mathscr{R}_{\alpha_n}^{-1}(\hat{\mu} - \mu)\rangle + \langle\mu, \mathscr{R}_{\alpha_n}^{-1}\mu\rangle. \tag{10}$$

For the first term on the right we compute

$$\begin{aligned}
|\langle\mu, (\hat{\mathscr{R}}_{\alpha_n}^{-1} - \mathscr{R}_{\alpha_n}^{-1})\hat{\mu}\rangle| &\leq \|\mu\|\|\hat{\mathscr{R}}_{\alpha_n}^{-1} - \mathscr{R}_{\alpha_n}^{-1}\|_\infty\|\hat{\mu}\| \\
&= \|\mu\|\|\hat{\mathscr{R}}_{\alpha_n}^{-1}(\hat{\mathscr{R}}_{\alpha_n} - \mathscr{R}_{\alpha_n})\mathscr{R}_{\alpha_n}^{-1}\|_\infty\|\hat{\mu}\| \\
&\leq \|\mu\|\|\hat{\mathscr{R}}_{\alpha_n}^{-1}\|_\infty\|\hat{\mathscr{R}}_{\alpha_n} - \mathscr{R}_{\alpha_n}\|_\infty\|\mathscr{R}_{\alpha_n}^{-1}\|_\infty\|\hat{\mu}\| \\
&\leq \alpha_n^{-2} O_P(n^{-1/2}),
\end{aligned}$$

since $\|\hat{\mathscr{R}}_{\alpha_n}^{-1}\|_\infty \leq \alpha_n^{-1}$, $\|\mathscr{R}_{\alpha_n}^{-1}\|_\infty \leq \alpha_n^{-1}$, $\|\hat{\mu}\| = O_P(1)$ and $\|\hat{\mathscr{R}}_{\alpha_n} - \mathscr{R}_{\alpha_n}\|_\infty = \|\hat{\mathscr{R}} - \mathscr{R}\|_\infty = O_P((n_1 + n_2)^{-1/2})$ (see Proposition 1 of Kraus, 2015). For the second term on the right side of (10) we obtain

$$|\langle \mu, \mathscr{R}_{\alpha_n}^{-1}(\hat{\mu} - \mu)\rangle| \leq \|\mu\| \|\mathscr{R}_{\alpha_n}^{-1}\|_\infty \|\hat{\mu} - \mu\| \leq \alpha_n^{-1} O_P(n^{-1/2}).$$

Finally the last term in (10) converges to $\|\mathscr{R}^{-1/2}\mu\|^2$ (finite or infinite) by the monotone convergence theorem like in the proof of Proposition 3.

The quantity in the denominator in (9) can be rewritten as

$$\langle \hat{\psi}_{\alpha_n}^{\text{Ridge}}, \mathscr{R}\hat{\psi}_{\alpha_n}^{\text{Ridge}}\rangle = \langle \hat{\psi}_{\alpha_n}^{\text{Ridge}} - \psi_{\alpha_n}^{\text{Ridge}}, \mathscr{R}\hat{\psi}_{\alpha_n}^{\text{Ridge}}\rangle + \langle \psi_{\alpha_n}^{\text{Ridge}}, \mathscr{R}(\hat{\psi}_{\alpha_n}^{\text{Ridge}} - \psi_{\alpha_n}^{\text{Ridge}})\rangle \\ + \langle \psi_{\alpha_n}^{\text{Ridge}}, \mathscr{R}\psi_{\alpha_n}^{\text{Ridge}}\rangle. \quad (11)$$

The first term on the right is

$$\langle \hat{\psi}_{\alpha_n}^{\text{Ridge}} - \psi_{\alpha_n}^{\text{Ridge}}, \mathscr{R}\hat{\psi}_{\alpha_n}^{\text{Ridge}}\rangle = \langle \hat{\mathscr{R}}_{\alpha_n}^{-1}\hat{\mu} - \mathscr{R}_{\alpha_n}^{-1}\mu, \mathscr{R}\hat{\mathscr{R}}_{\alpha_n}^{-1}\hat{\mu}\rangle \\ = \langle \mathscr{R}_{\alpha_n}^{-1}(\mathscr{R}_{\alpha_n} - \hat{\mathscr{R}}_{\alpha_n})\hat{\mathscr{R}}_{\alpha_n}^{-1}\hat{\mu}, \mathscr{R}\hat{\mathscr{R}}_{\alpha_n}^{-1}\hat{\mu}\rangle + \langle \mathscr{R}_{\alpha_n}^{-1}(\hat{\mu} - \mu), \mathscr{R}\hat{\mathscr{R}}_{\alpha_n}^{-1}\hat{\mu}\rangle.$$

Here we compute for the first summand

$$|\langle \mathscr{R}_{\alpha_n}^{-1}(\mathscr{R}_{\alpha_n} - \hat{\mathscr{R}}_{\alpha_n})\hat{\mathscr{R}}_{\alpha_n}^{-1}\hat{\mu}, \mathscr{R}\hat{\mathscr{R}}_{\alpha_n}^{-1}\hat{\mu}\rangle| \leq \|\hat{\mu}\|^2 \|\hat{\mathscr{R}}_{\alpha_n}^{-1}\|_\infty^2 \|\mathscr{R}\mathscr{R}_{\alpha_n}^{-1}\|_\infty \|\hat{\mathscr{R}} - \mathscr{R}\|_\infty \\ \leq \alpha_n^{-2} O_P(n^{-1/2})$$

using properties mentioned previously and $\|\mathscr{R}\mathscr{R}_{\alpha_n}^{-1}\|_\infty \leq 1$ and for the second summand

$$|\langle \mathscr{R}_{\alpha_n}^{-1}(\hat{\mu} - \mu), \mathscr{R}\hat{\mathscr{R}}_{\alpha_n}^{-1}\hat{\mu}\rangle| \leq \|\mathscr{R}\mathscr{R}_{\alpha_n}^{-1}\|_\infty \|\hat{\mathscr{R}}_{\alpha_n}^{-1}\|_\infty \|\hat{\mu} - \mu\| \leq \alpha_n^{-1} O_P(n^{-1/2}).$$

Putting these results together we see that the absolute value of the first term on the right-hand side in (11) is dominated by $\alpha_n^{-2} O_P(n^{-1/2})$. Next, the second term on the right in (11) can be analyzed like the first two terms on the right in (10) with $\mathscr{R}\mathscr{R}_{\alpha_n}^{-1}\mu$ in place of $\mu$. This way we bound its absolute value from above by $\alpha_n^{-2} O_P(n^{-1/2})$. The last term in (11) converges to $\|\mathscr{R}^{-1/2}\mu\|^2$ like in the proof of Proposition 3 (for both finite and infinite limiting value).

Now if $\|\mathscr{R}^{-1/2}\mu\| < \infty$, the proof is complete. Otherwise we can proceed like in the proof of Proposition 3.

## B   Details of the simulation settings

The mean difference $\mu(t)$, $t \in [0,1]$ in simulations in Subsections 5.1 and 5.2 takes the form (i) $ct$, (ii) $c(t-0.5)^2$, (iii) $c(t-0.5)^3$, (iv) $c\sin(20t)$, (v) $c\varphi_1(t)$, (vi) $c\varphi_{10}(t)$, (vii) $cb(t;5,5)$, (viii) $cb(t;2,6)$, where $b(t;\alpha,\beta) = t^{\alpha-1}(1-t)^{\beta-1}$. In Subsection 5.3, $\mu(t)$ takes the form (i) $cb(t;2,6)$, (ii) $cb(t;5,5)$, (iii) $cb(t;6,2)$. The parameter $c$ is selected in each case to yield a reasonable misclassification rate.

# References

Baíllo, A., Cuevas, A., and Cuesta-Albertos, J. A. (2011a). Supervised classification for a family of Gaussian functional models. *Scandinavian Journal of Statistics*, 38(3):480–498.

Baíllo, A., Cuevas, A., and Fraiman, R. (2011b). Classification methods for functional data. In *The Oxford Handbook of Functional Data Analysis*, pages 259–297. Oxford University Press, Oxford.

Berrendero, J. R., Cuevas, A., and Torrecilla, J. L. (2017). On the use of reproducing kernel Hilbert spaces in functional classification. *Journal of the American Statistical Association*. To appear.

Blanchard, G. and Krämer, N. (2010). Kernel partial least squares is universally consistent. In *Proceedings of the Thirteenth International Conference on Artificial Intelligence and Statistics*, pages 57–64.

Bosq, D. (2000). *Linear Processes in Function Spaces*. Springer, New York.

Bro, R. and Eldén, L. (2009). PLS works. *Journal of Chemometrics*, 23(2):69–71.

Bugni, F. A. (2012). Specification test for missing functional data. *Econometric Theory*, 28(5):959–1002.

Cardot, H., Ferraty, F., and Sarda, P. (1999). Functional linear model. *Statistics & Probability Letters*, 45(1):11–22.

Cuesta-Albertos, J. A., del Barrio, E., Fraiman, R., and Matrán, C. (2007). The random projection method in goodness of fit for functional data. *Computational Statistics & Data Analysis*, 51(10):4814–4831.

Cuevas, A. (2014). A partial overview of the theory of statistics with functional data. *Journal of Statistical Planning and Inference*, 147:1–23.

De Jong, S. (1993). SIMPLS: an alternative approach to partial least squares regression. *Chemometrics and Intelligent Laboratory Systems*, 18(3):251–263.

Delaigle, A. and Hall, P. (2012a). Achieving near perfect classification for functional data. *Journal of the Royal Statistical Society. Series B. Statistical Methodology*, 74(2):267–286.

Delaigle, A. and Hall, P. (2012b). Methodology and theory for partial least squares applied to functional data. *The Annals of Statistics*, 40(1):322–352.

Delaigle, A. and Hall, P. (2013). Classification using censored functional data. *Journal of the American Statistical Association*, 108(504):1269–1283.

Delaigle, A. and Hall, P. (2016). Approximating fragmented functional data by segments of Markov chains. *Biometrika*, 103(4):779–799.

Delaigle, A., Hall, P., and Bathia, N. (2012). Componentwise classification and clustering of functional data. *Biometrika*, 99(2):299–313.

Febrero-Bande, M., Galeano, P., and González-Manteiga, W. (2017). Functional principal component regression and functional partial least-squares regression: An overview and a comparative study. *International Statistical Review*, 85(1):61–83.

Ferraty, F., Hall, P., and Vieu, P. (2010). Most-predictive design points for functional data predictors. *Biometrika*, 97(4):807–824.

Goldberg, Y., Ritov, Y., and Mandelbaum, A. (2014). Predicting the continuation of a function with applications to call center data. *J. Statist. Plann. Inference*, 147:53–65.

Gromenko, O., Kokoszka, P., and Sojka, J. (2017). Evaluation of the cooling trend in the ionosphere using functional regression with incomplete curves. *The Annals of Applied Statistics*, 11(2):898–918.

Hastie, T., Tibshirani, R., and Friedman, J. (2009). *The Elements of Statistical Learning*. Springer, New York.

Horváth, L. and Kokoszka, P. (2012). *Inference for Functional Data with Applications*. Springer, New York.

Hsing, T. and Eubank, R. (2015). *Theoretical Foundations of Functional Data Analysis, with an Introduction to Linear Operators*. John Wiley & Sons.

Kraus, D. (2015). Components and completion of partially observed functional data. *Journal of the Royal Statistical Society. Series B. Statistical Methodology*, 77(4):777–801.

Liebl, D. (2013). Modeling and forecasting electricity spot prices: a functional data perspective. *The Annals of Applied Statistics*, 7(3):1562–1592.

Lingjærde, O. C. and Christophersen, N. (2000). Shrinkage structure of partial least squares. *Scandinavian Journal of Statistics. Theory and Applications*, 27(3):459–473.

Phatak, A. and de Hoog, F. (2002). Exploiting the connection between PLS, Lanczos methods and conjugate gradients: alternative proofs of some properties of PLS. *Journal of Chemometrics*, 16(7):361–367.

Ramsay, J. O. and Silverman, B. W. (2005). *Functional Data Analysis*. Springer, New York.

Sangalli, L. M., Secchi, P., and Vantini, S. (2014a). Analysis of AneuRisk65 data: $k$-mean alignment. *Electronic Journal of Statistics*, 8(2):1891–1904.

Sangalli, L. M., Secchi, P., and Vantini, S. (2014b). AneuRisk65: A dataset of three-dimensional cerebral vascular geometries. *Electronic Journal of Statistics*, 8(2):1879–1890.

Sangalli, L. M., Secchi, P., Vantini, S., and Veneziani, A. (2009a). A case study in exploratory functional data analysis: geometrical features of the internal carotid artery. *Journal of the American Statistical Association*, 104(485):37–48.

Sangalli, L. M., Secchi, P., Vantini, S., and Veneziani, A. (2009b). Efficient estimation of three-dimensional curves and their derivatives by free-knot regression splines, applied to the analysis of inner carotid artery centrelines. *Journal of the Royal Statistical Society. Series C. Applied Statistics*, 58(3):285–306.